\documentclass[runningheads]{llncs}
\usepackage{graphicx}
\usepackage{mathtools}
\usepackage{amsmath}
\usepackage{bbding} 
\usepackage{enumerate}
\usepackage{authblk}
\usepackage{url}

\urldef{\mailsZhang}\path|hanxiao.zhang18@imperial.ac.uk|
\urldef{\mailsLi}\path|jingxiong.li@se18.qmul.ac.uk|
\urldef{\mailsWang}\path|wangyaqi@hdu.edu.cn|

\begin{document}
\title{DDU-Nets: Distributed Dense Model for 3D MRI Brain Tumor Segmentation}
%
%

\author{Hanxiao Zhang\inst{1, 2}, Jingxiong Li\inst{3}, Mali Shen\inst{2}, Yaqi Wang\inst{4},\authorcr and 
Guang-Zhong Yang\inst{1, 2}}
%
%
\authorrunning {Hanxiao Zhang et al.}
\institute{The Institute of Medical Robotics, Shanghai Jiao Tong University, China \\ \and The Hamlyn Centre for Robotic Surgery, Imperial College London, UK  \\ \mailsZhang \and
Queen Mary University of London, London E1 4NS, UK\\  \and
Hangzhou Dianzi University, Hangzhou 310018, China\\
} 

\maketitle              
\begin{abstract}

Segmentation of brain tumors and their subregions remains a challenging task due to their weak features and deformable shapes. In this paper, three patterns (cross-skip, skip-1 and skip-2) of distributed dense connections (DDCs) are proposed to enhance feature reuse and propagation of CNNs by constructing tunnels between key layers of the network. For better detecting and segmenting brain tumors from multi-modal 3D MR images, CNN-based models embedded with DDCs (DDU-Nets) are trained efficiently from pixel to pixel with a limited number of parameters. Postprocessing is then applied to refine the segmentation results by reducing the false-positive samples. The proposed method is evaluated on the BraTS 2019 dataset with results demonstrating the effectiveness of the DDU-Nets while requiring less computational cost.


\end{abstract}
\section{Introduction}
Gliomas are a kind of brain tumor developed from glial cells. It is one of the most threatening brain tumors as more than 40 percent of all tumors befall are malignant \cite{mamelak2007targeted}. As a result, it is necessary to develop an accurate segmentation model for quantitative assessment of brain tumors, assisting early diagnosis and treatment planning. However, because of the diverse characteristics of tumor cells, reliable tumor segmentation remains a challenging task.

Focusing on the evaluation of state-of-the-art brain tumor segmentation methods, the annual Brain Tumor Segmentation Challenge (BraTS) provides datasets of brain magnetic resonance imaging (MRI) scans collected from multiple institutions \cite{2bakas2017segmentation}\cite{bakas2017segmentation}\cite{Bakas2017AdvancingTC}\cite{bakas2018identifying}\cite{menze2014multimodal}. The datasets include annotated MRI scans of low grade gliomas (LGG) and high grade glioblastomas (GMM/HGG), acquired under standard clinical conditions with different equipment and protocols. For each case, four 3D MRI modalities are provided consisting of a native T1-weighted (T1), a post-contrast T1-weighted scan (T1Gd), a native T2-weighted scan (T2) and a T2 Fluid Attenuated Inversion Recovery (T2-FLAIR) scan. Each tumor is divided into 3 subregions for evaluation, which are enhancing tumor (ET), tumor core (TC) and whole tumor (WT), referred as complete tumor region extent \cite{bakas2018identifying}. All labels are evaluated manually by professional raters and approved by internationally recognized expert neuroradiologists.

Recently, convolutional neural networks (CNNs) with encoder-decoder structure have demonstrated their ability in segmenting biomedical images \cite{cao2018improve}\cite{dong2017automatic}\cite{li2018h}\cite{ronneberger2015u}, where the encoder down-samples and extracts features from the input data while the decoder rebuilds segmentation of the targets. As a result, methods based on CNNs are popular for brain tumor segmentation. In 2018, Myronenko \cite{myronenko20183d}, who won the 1st prize in BraTS18, proposed their ResNet based 3D encoder-decoder model and achieved the highest accuracy and robustness among others. Isensee et al. \cite{isensee2018no} also shown that well-trained U-Net without much modification could achieve competitive segmentation accuracy. 

Current challenges of CNN-based methods include false predictions caused by weak features of tumors, gradient vanishing and overfitting problems when training on deep CNNs, slow training speed due to a large amount of training data, and low accuracy caused by false-positive predictions.
To deal with these problems, we propose in this paper several patterns of distributed dense connections (DDCs), which reuse features in different strategies. CNN-based models with DDCs are designed to automatically segment brain tumor targets. In addition, postprocessing is applied to reduce the false-positives for more accurate delineation.



\section{Methods}
\subsection{Distributed Dense Connections (DDCs)}
\label{Sec:DDC}

Although deeper CNNs could reach a better performance than that of shallow ones, the problem of gradient vanishing can have a negative impact on network capacity and efficiency. It has been shown that this can be alleviated by shortcut connections between contextual layers \cite{he2016deep}\cite{huang2017densely}\cite{larsson2016fractalnet}\cite{pleiss2017memory}.

The concept of residual learning network is used by ResNet \cite{he2016deep} to address the degradation problem, which uses shortcut connections to skip one or more layers by summation, providing implicit deep supervision. DenseNet \cite{huang2017densely} proposes dense blocks with more shortcut connections, combining the feature maps of all the preceding layers as the input of the subsequent layer using a more efficient concatenation strategy. In practice, DenseNet provides better performance but consumes more GPU memory as the number of input channels grows dramatically towards deeper layers. The structure of DenseNet only allows dense connections being operated within each dense block and no shortcut connections are operated between dense blocks. It shows that GPU memory consumption can be reduced at the expense of training time by introducing an implementation strategy of shared memory allocations for storing feature maps \cite{pleiss2017memory}. Different from this memory-saving strategy, we aim to improve our network efficiency, while performing a better accuracy with much fewer dense connections.


We note that the gradient is unlikely to vanish very quickly in a few layers, so there is no need to relearn redundant features right after the preceding layers. The feature reuse can also be strengthened when the early feature maps are recalled by a reasonable skip distance.

Consequently, we propose a novel densely connected unit called distributed dense connections (DDCs), which only transmit features between critical intermediate layers without settled blocks. 
This approach reduces the total number of dense connections and extends the radiation scope of early features to the deeper layers, thus reducing the number of parameters and enhancing the global integration of information flow.
The identity function of a neural network using distributed dense connections can be expressed as:
\begin{equation}
    x_{n}\:=\:F_{n}\:([x_{n_{i}},\:x_{n_{ii}},\:\cdots,\:x_{n_{i\cdots ii}}])\label{eq3.4}
\end{equation}
where $F_{n}(\cdot)$ is non-linear transformation after each layer, $n$ represents the $n^{th}$ layer. The output of the $n^{th}$ layer is represented as $x_{n}$. $[x_{n_{i}},\:x_{n_{ii}},\:\cdots,\:x_{n_{i\cdots ii}}]$ refers to the concatenation of the feature maps produced by the chosen layers $n_{i},\:n_{ii},\:\cdots,\:n_{i\cdots ii}$.


\begin{figure}[htb]
\centering
\includegraphics[width = 100 mm]{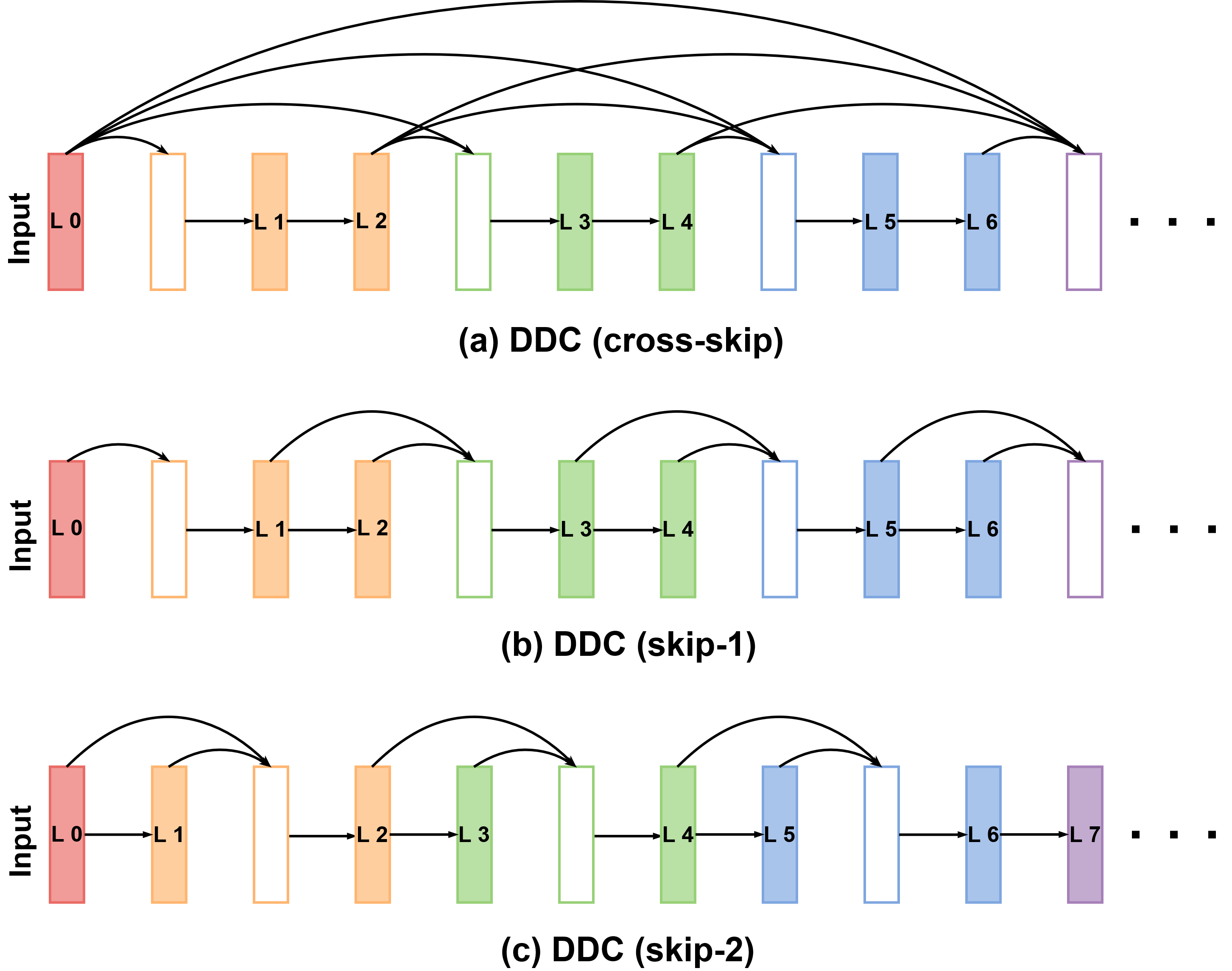}
\caption{Three architectures of distributed dense connections. The boxes with the same color denote hidden layers with the same size, known as the size-block. White boxes with colored borders mean concatenation operation for chosen layers. (a) stands for cross-skip, which reuses feature maps from the last layer of each size-block; (b) represents skip-1, which reuses the last 2 layers of preceding size-block; (c) represents skip-2, which reuses the last layer of preceding size-block and the first layer of the current block.}
\label{fig1} 
\end{figure}

To match different sizes of feature maps, we consider three solutions: (a) Downsample the upper layer by performing $2\times2$ convolutions with the stride of 2; (b) $3\times3$ dilated convolutions with dilation 2 may increase the receptive field of feature maps when performing downsampling by the stride of 2; (c) Pooling is a simple way to halve the size of feature maps, which generates no additional parameter. In the network that is not deep enough, we recommend (average) pooling to extract different implicit features from former layers before passing to the back layers which may contain max pooling features.

Three patterns of distributed dense connections are designed in Figure \ref{fig1}, varying in terms of choosing key layers and the methods for transmitting feature maps. In (a), DDC (cross-skip), each concatenation input consists of features from the final layer of all size-blocks. Due to the global transmission of chosen features, each size-block could reuse features from all the preceding blocks.
For (b), each concatenation input includes features only chosen from the last two layers of preceding size-block where the information flow does not spread globally. We name this pattern as 'skip-1'. (c) reuses the features from 2 size-blocks. Each concatenation input consists of features chosen from the first layer of present size-block and the last layer of the preceding size block.

\subsection{DDU-Net Architectures}

\begin{figure}[!hbt]
\centering
\includegraphics[width = 100 mm]{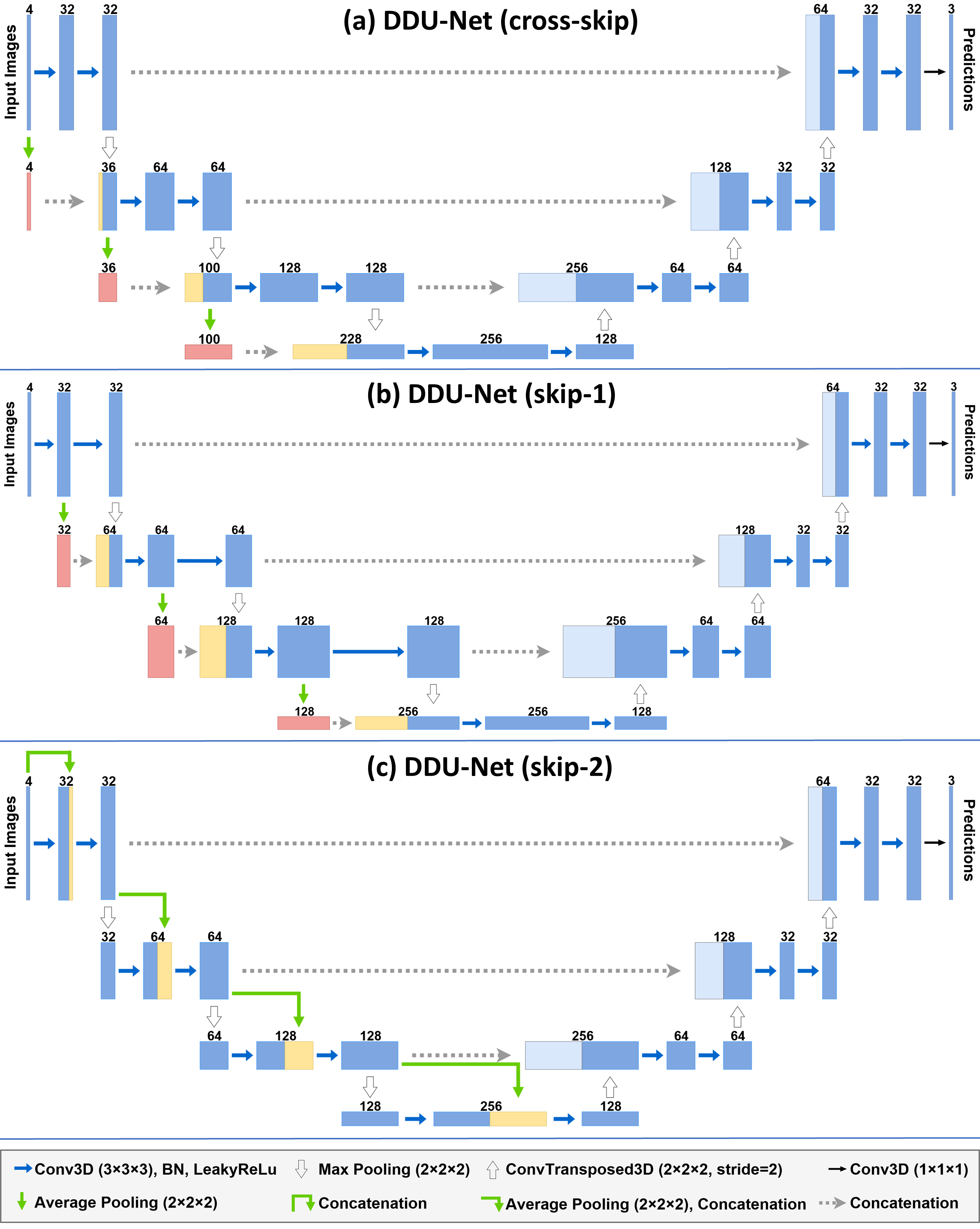}
\caption{DDU-Nets with different patterns of DDCs. (a), (b) and (c) represent DDU-Nets with cross-skip, skip-1 and skip-2 respectively. All the feature maps and translations are based on 3D volumes. Copied feature maps and concatenated feature maps can be distinguished by different colors of boxes. Operations are represented by arrows.}
\label{fig2} 
\end{figure}

Inspired by the encoder-decoder architecture, which is widely used for biomedical image segmentation, We modified U-Net \cite{ronneberger2015u} respectively by adding the above three patterns of distributed dense connections between each neighboring resolution stages in the encoder path. These proposed networks are named as DDU-Nets (distributed dense U-Nets), as shown in Figure \ref{fig2}.

The networks in Figure \ref{fig2} inherit the encoder-decoder architecture with 4 resolution stages (levels) operating with different sizes and channels of layers. Every stage in the encoder path consists of two convolutional layers with $3\times3\times3$ kernels applied by 1 stride and 1 padding, each followed by a LeakyReLU ($alpha = 0.2$) and a 3D batch normalization. Max pooling with the stride of 2 is applied at each end of the encoder stage to downsize the feature maps. In the decoder path, each stage has an up-sampling operation and two convolutions each followed by a LeakyReLU and a batch normalization. In order to achieve the pixel-to-pixel localization, the feature maps at each end of the encoder stage are concatenated to the beginning of the corresponding decoder stage, which provides the high-resolution features to the decoder path. At the final layer, a $1\times1\times1$ convolution is used to produce the output with the required numbers of classes and the same image size as the input data.

At the encoder side of the network, we apply distributed dense connection to bridge over features between stages. 
For cross-skip pattern shown in Figure \ref{fig2} (a), the first layer of the upper stage in the encoder path is down-sampled by average pooling before concatenation, aiming to match the size of the first layer of the lower stage, which is also the output of the upper stage after max pooling. 
The input of each stage has direct access to all the previous representative feature maps on a global scale, thus enhancing the feature reuse and propagation with less redundant connections. 
Similarly, the DDU-Nets of skip-1 (b) and skip-2 (c) in Figure \ref{fig2} follow the patterns of their own distributed dense connections mentioned in Section \ref{Sec:DDC}.
With the application of distributed dense connections, we empirically halve the feature channels of each stage comparing to the traditional U-Net \cite{ronneberger2015u}. Experiments (see Table \ref{tab1}) show that the proposed network architectures can effectively improve the performance of brain tumor segmentation tasks.   


\subsection{Loss Function}

Our loss function includes two parts: the average of Dice loss and L2 regularization which are shown in equation \eqref{eq1} and \eqref{eq2}:

\begin{equation}
    L_{Dice} = \sum_{class}(1 - DICE)\label{eq1}
\end{equation}

\begin{equation}
    L = L_{Dice(mean)} + 0.01 L_{L2(total)}\label{eq2}
\end{equation}

 To accurately represent the loss of inference, the Dice coefficient is used to represent the loss function, which is a frequently used measurement for pixel-wise segmentation tasks. One of the main challenges of brain tumor segmentation is the imbalance of each subregion. We try to reduce the impact by calculating the average value of three Dice loss functions for three output channels (predictions for each subregion), instead of calculating the Dice loss for the entire predictions directly. For the regularization part, L2 loss displays on the entire predictions and is assigned a hyper-parameter weight to prevent overfitting. 

\subsection{Training Configuration}

BraTS 2019 dataset contains non-standardized 3D images with the size of $250\times250\times155$. Since the data is from different institutes, the value could vary due to different MRI machines or configurations. To ease these impacts and reduce the initial bias caused by the variations of cases, z-score standardization transform is applied to each of the four image modalities before concatenating them into an input with four channels. Then we reassigned four different labels (label 0, label 1, label 2, label 4) of ground truth into three combined subregions (see Figure 3), representing enhancing tumor (label 4), tumor core (label 1 + label 4) and whole tumor (label 1 + label 2 + label 4), respectively, to optimize the segmentation accuracy for each region independently in the model. Therefore, the final layer of the network has three channels for the three subregions and we use sigmoid instead of softmax to output the segmentation predictions.

The network is implemented in Pytorch and trained using Adam optimizer with the learning rate of 3e-4. We run our operations parallelly on two GPUs (GeForce GTX 1080 Ti: 11G; TITAN Xp: 12G). In order to fit the capability of our network within GPU memory limits, we cropped all the data into a size of $192\times192\times128$, and then extracted three smaller overlapping volumes ($192\times192\times64$) by the stride of 32 in the third dimension. Partitioning images with overlapping area served as a type of data augmentation that ensures the seamless cohesion of separated small volumes, preventing information loss due to cropping. The batch size is 4 and trained the network for 100 epochs (335 cases for each epoch), taking 16 hours in total.


\subsection{Postprocessing}

In some of the low grade gliomas (LGG) cases, there is no existence of enhancing tumor while the model may infer as existing, causing large error in Dice coefficient. Thus, if the number of voxels classified as the segmented enhancing tumor (ET) is less than 300 in a single case, those voxels are regarded as false-positive for ET (label 4) and replaced with the label of necrotic and non-enhancing tumor parts (label 1). Some independent small volumes disconnected with the largest tumor area are removed by connected component processing. If the voxel number of each small component is less than 30 percent of the total number of predicted class, those components were re-labeled as background.

\section{Results}

\begin{figure}[htb]
\centering
\includegraphics[width = 90 mm]{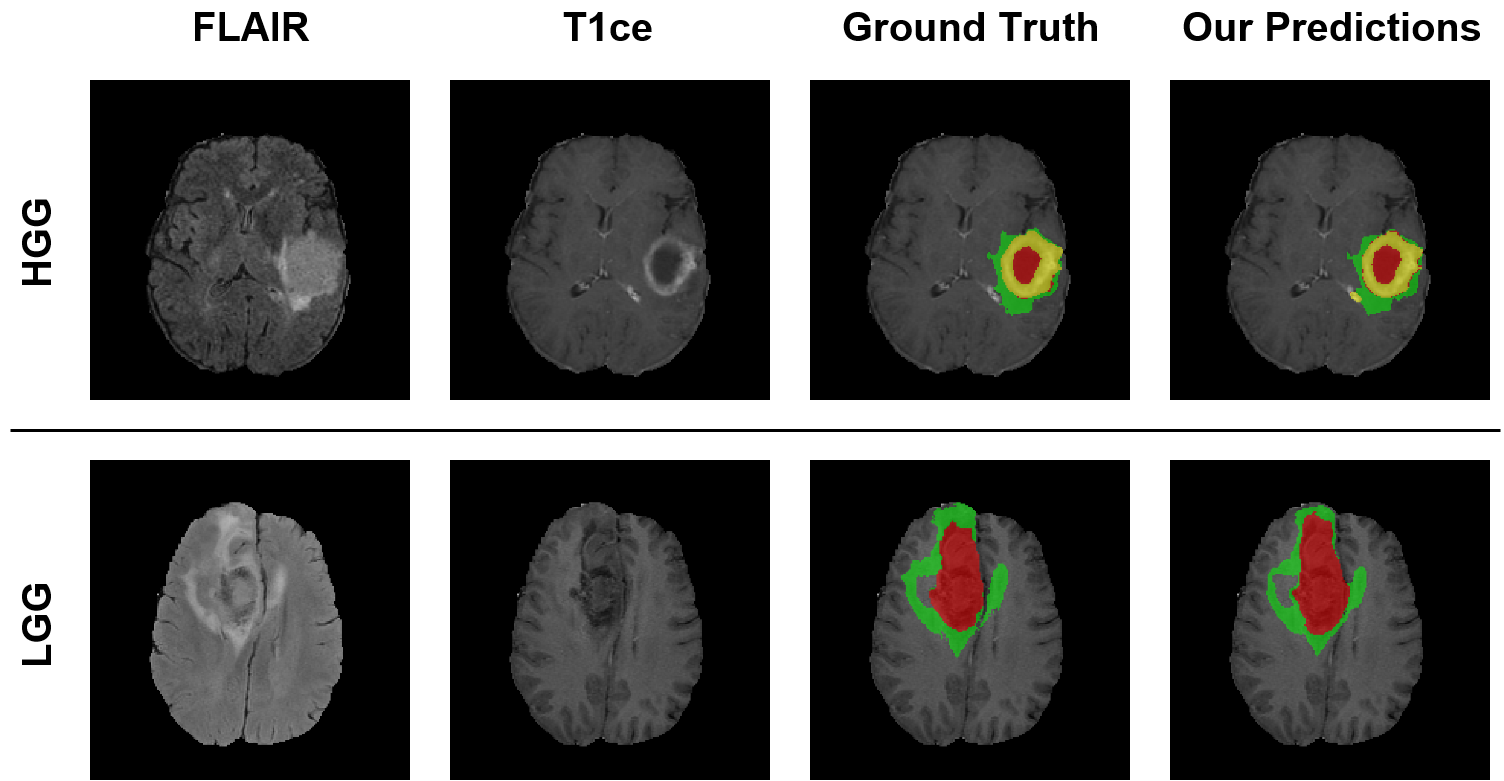}
\caption{Visualization results of two examples using DDU-Net (cross-skip) with corresponding FLAIR slices, T1ce slices and ground truth on BraTS 2019 Training dataset. Yellow: enhancing tumor (label 4); Red: necrotic and non-enhancing tumor core (label 1); Green: peritumoral edema (label 2).}
\label{fig3} 
\end{figure}

The proposed models are trained on the BraTS 2019 training dataset (335 cases) and initially evaluated on the validation dataset (125 cases). All the predicted results after reconstruction and post-processing are uploaded for the generalizability assessments by CBICA's Image Processing Portal (IPP). Example segmentation results are shown in Figure \ref{fig3}.

The performance of proposed models is evaluated and compared with baselines (U-Net and DU-net) by Dice score, Sensitivity, Specificity and Hausdorff distance (95\%) for all subregions. 
Four metric average results of three subregions on the BraTS 2019 validation dataset are presented in Table \ref{tab1}.

\begin{table*}[]
\centering
\caption{Models evaluated (mean value) on BraTS 2019 validation dataset. ET, WT and TC denote enhancing tumor, whole tumor and tumor core, respectively. P, cs, s1, s2 stand for postprocessing, cross-skip, skip-1, skip-2. DU-Net (dense U-Net) is an integrated model of the traditional dense connection and U-Net.}
\label{tab1}
\resizebox{\textwidth}{!}{
\begin{tabular}{l|c|c|c|c|c|c|c|c|c|ccc}
\hline \hline
 & \multicolumn{3}{c|}{Dice score} & \multicolumn{3}{c|}{Sensitivity} & \multicolumn{3}{c|}{Specificity} & \multicolumn{3}{c}{95 Hausdorff} \\ \hline
Models & ET & WT & TC & ET & WT & TC & ET & WT & TC & \multicolumn{1}{c|}{ET} & \multicolumn{1}{c|}{WT} & TC \\ \hline \hline
U-Net & \multicolumn{1}{l|}{0.744} & \multicolumn{1}{l|}{0.893} & \multicolumn{1}{l|}{0.765} & \multicolumn{1}{l|}{0.740} & \multicolumn{1}{l|}{0.883} & \multicolumn{1}{l|}{0.763} & \multicolumn{1}{l|}{0.997} & \multicolumn{1}{l|}{\textbf{0.995}} & \multicolumn{1}{l|}{\textbf{0.997}} & \multicolumn{1}{l|}{4.712} & \multicolumn{1}{l|}{6.279} & \multicolumn{1}{l}{8.146} \\
U-Net+P & 0.776 & 0.894 & 0.768 & 0.795 & 0.883 & 0.770 & \textbf{0.998} & \textbf{0.995} & \textbf{0.997} & \multicolumn{1}{c|}{3.470} & \multicolumn{1}{c|}{4.968} & 7.882 \\
DU-Net+P & 0.767 & 0.887 & 0.773 & 0.758 & 0.883 & 0.763 & \textbf{0.998} & \textbf{0.995} & \textbf{0.997} & \multicolumn{1}{c|}{3.843} & \multicolumn{1}{c|}{5.908} & 7.989 \\ \hline \hline
DDU-Net(cs)+P & 0.780 & \textbf{0.898} & 0.793 & 0.791 & 0.903 & 0.808 & \textbf{0.998} & 0.994 & 0.996 & \multicolumn{1}{c|}{\textbf{3.376}} & \multicolumn{1}{c|}{\textbf{4.874}} & 8.013 \\
DDU-Net(s1)+P & 0.765 & \textbf{0.898} & 0.793 & 0.787 & \textbf{0.905} & \textbf{0.820} & \textbf{0.998} & 0.994 & 0.996 & \multicolumn{1}{c|}{4.058} & \multicolumn{1}{c|}{5.225} & 8.127 \\
DDU-Net(s2)+P & \textbf{0.784} & 0.897 & \textbf{0.794} & \textbf{0.804} & 0.888 & 0.791 & \textbf{0.998} & \textbf{0.995} & \textbf{0.997} & \multicolumn{1}{c|}{4.099} & \multicolumn{1}{c|}{4.950} & \textbf{7.399} \\ \hline \hline
\end{tabular}}
\end{table*}



To illustrate the effectiveness of postprocessing method, experiments on U-Net have been conducted. The results are shown in Table \ref{tab1} (the $1^{st}$ and the $2^{nd}$ row). Comparing their performance on metrics, the effectiveness of the method is evident, especially for enhancing tumor.
To explore the effects of DDCs, detailed experiments have been operated on each DDU-Net. Compared with U-Net baseline and dense U-net (DU-Net) using dense connection to bridge over features, the results show that the performance of the DDU-Nets surpasses that of baselines on Dice score and sensitivity in most of the subregions. Although DU-Net possesses the highest feature reuse rate, the redundant architecture makes it difficult to achieve better performance than that of the DDU-Nets, which reuse fewer features but achieve great improvement.
For each model, Specificity has no obvious difference.

Within the DDU-Net models, different architectures dictate the relative performance that each metric of subregions displays.
Cross-skip excels at edge characterization due to the global feature reuse of the localization information provided by original images.
Skip-1 has the advantage in terms of Sensitivity at the expense of low Dice score in enhancing tumor.
Skip-2 achieves a good result in Dice score, with a low feature reuse rate because of the specific design of DDU-Net (skip-2) that preserving neighborhood information in an adequate skip distance (2 stages). Overall, DDU-Nets with cross-skip pattern can be considered to acquire the best comprehensive performance with an easy model deployment.


\begin{table}[]
\caption{Performances (mean value) of DDU-Net (cross-skip) with different architectures on BraTS 2019 validation dataset.}
\label{tab2}
\resizebox{\textwidth}{!}{
\begin{tabular}{l|c|c|c|c|c|c|c|c|c|ccc}
\hline
 & \multicolumn{3}{c|}{Dice} & \multicolumn{3}{c|}{Sensitivity} & \multicolumn{3}{c|}{Specificity} & \multicolumn{3}{c}{95 Hausdorff} \\ \hline
Method & ET & WT & TC & ET & WT & TC & ET & WT & TC & \multicolumn{1}{c|}{ET} & \multicolumn{1}{c|}{WT} & TC \\ \hline
Stage 5 & \textbf{0.782} & 0.896 & 0.784 & \textbf{0.812} & 0.905 & 0.799 & 0.998 & \textbf{0.994} & 0.996 & \multicolumn{1}{c|}{4.392} & \multicolumn{1}{c|}{5.616} & 8.023 \\
Ch 64 & 0.772 & 0.873 & 0.786 & 0.780 & \textbf{0.935} & \textbf{0.817} & 0.998 & 0.988 & 0.995 & \multicolumn{1}{c|}{4.699} & \multicolumn{1}{c|}{7.984} & 8.260 \\
Vol 128 & 0.674 & 0.862 & 0.640 & 0.661 & 0.872 & 0.617 & \textbf{0.999} & 0.993 & \textbf{0.997} & \multicolumn{1}{c|}{6.370} & \multicolumn{1}{c|}{6.379} & 10.846 \\
\textbf{Final} & 0.780 & \textbf{0.898} & \textbf{0.793} & 0.791 & 0.903 & 0.808 & 0.998 & \textbf{0.994} & 0.996 & \multicolumn{1}{c|}{\textbf{3.376}} & \multicolumn{1}{c|}{\textbf{4.874}} & \textbf{8.013} \\ \hline
\end{tabular}}
\end{table}

\par Apart from the approaches mentioned above, other potential architectures and postprocessing methods also deployed during our experiments. As Table \ref{tab2} shown, we attempted to allocate DDU-Net (cross-skip) with more stages (5 stages) and feature channels (start with 64 channels in the first stage), but all led to worse results in general, which proves that the distributed dense connections can contribute better performance as well as decreasing in network depth and width. We tried to input with the larger volumes ($192\times192\times128$) without further cropping, but it didn't show better results as well. We also denied the opening operation solution used to denoise for the postprocessing which is replaced by connected component processing. Compared with those alternatives, the final proposed architecture achieved the best performance by balancing among the size of input data, the capability of network and the GPU memory consumption.

\begin{table}[]
\centering
\caption{Performance of DDU-Net (cross-skip) with postprocessing on BraTS 2019 testing dataset.}
\label{tab3}

\begin{tabular}{l|c|c|c|ccc}
\hline
& \multicolumn{3}{c|}{Dice} & \multicolumn{3}{c}{95 Hausdorff} \\ \cline{2-7} 
 & ET & WT & TC & \multicolumn{1}{c|}{ET} & \multicolumn{1}{c|}{WT} & TC \\ \hline
Mean & 0.804 & 0.876 & 0.821 & \multicolumn{1}{c|}{3.41} & \multicolumn{1}{c|}{7.054} & 6.774 \\
StdDev & 0.193 & 0.117 & 0.237 & \multicolumn{1}{c|}{7.44} & \multicolumn{1}{c|}{11.600} & 13.238 \\
Median & 0.846 & 0.917 & 0.908 & \multicolumn{1}{c|}{1.732} & \multicolumn{1}{c|}{3.535} & 3.000 \\
25quantile & 0.774 & 0.854 & 0.836 & \multicolumn{1}{l|}{1.414} & \multicolumn{1}{c|}{2.236} & 1.946 \\
75quantile & 0.916 & 0.943 & 0.948 & \multicolumn{1}{c|}{2.639} & \multicolumn{1}{c|}{5.916} & 5.745 \\ \hline
\end{tabular}
\end{table}

Table \ref{tab3} presents mean value, standard deviation, median, 25 and 75 quantiles of two metrics on BraTS 2019 testing dataset (166 cases). Due to limited one submission chance, we only evaluated DDU-Net (cross-skip) on testing dataset. The results demonstrate that the performance of this model is highly competitive, with mean Dice scores of 0.804, 0.876, and 0.821 for enhancing tumor, whole tumor and tumor core, respectively.

\section{Conclusion}

In conclusion, this paper has shown a new network structure for brain tumor segmentation. Three distributed dense connections (DDCs) have been proposed for generic CNNs to inherit features efficiently. DDU-Nets are built to verify the effectiveness of DDCs. Postprocessing is deployed to eliminate false-positive pixels. The results show that the DDU-Nets can segment 3D MR images effectively by allocating DDCs to key layers, among which DDU-Net with cross-skip pattern achieved the competitive performance.



\bibliographystyle{splncs04}

\bibliography{ref1}

\end{document}